\documentclass[3p]{elsarticle}

 \usepackage{color}
\usepackage{graphicx}			
\usepackage{dcolumn}			
\usepackage{bm}					
\usepackage{float}			
\usepackage{amsmath,amsfonts}	
\usepackage{url}					
\usepackage{setspace}		
\usepackage{lineno}   			
 \usepackage{color}
 \usepackage{color,soul}

\begin{document}

\begin{frontmatter}

\title{Mitigation of seismic waves: metabarriers and metafoundations bench tested}

\author[address1]{Andrea Colombi\corref{mycorrespondingauthor}}
\cortext[mycorrespondingauthor]{Corresponding author}
\ead{colombia@ethz.ch}
\author[address1]{Rachele Zaccherini}
\author[address2]{Antonio Palermo}

\address[address1]{Department of Civil, Environmental and Geomatic Engineering, ETH, Z\"urich, Switzerland}
\address[address2]{Dept. of Civil, Chemical, Environmental and Materials Engineering, Universit\`{a} di Bologna, Italy}



\begin{abstract}
The article analyses two potential metamaterial designs, the metafoundation and the metabarrier, capable to attenuate seismic waves on buildings or structural components in a frequency band between 3.5 to 8 Hz. The metafoundation serves the dual purpose of reducing the seismic response and supporting the superstructure. Conversely the metabarrier surrounds and shields the structure from incoming waves. The two solutions are based on a cell layout of local resonators whose dynamic properties are tuned using finite element simulations combined with Bloch-Floquet boundary conditions.  To enlarge the attenuation band, a graded design where the resonant frequency of each cell varies spatially is employed. If appropriately enlarged or reduced, the metamaterial designs could be used to attenuate lower frequency seismic waves or groundborne vibrations respectively.
A sensitivity analysis over various design parameters including size, number of resonators, soil type and source directivity, carried out by computing full 3D numerical simulations in time domain for horizontal shear waves is proposed. Overall, the metamaterial solutions discussed here can reduce the spectral amplification of the superstructure between approx. 15 to 70\% depending on several parameters including the metastructure size and the properties of the soil. Pitfalls and advantages of each configuration are discussed in detail. The role of damping, crucial to avoid multiple resonant coupling, and the analogies between graded metamaterials and tuned mass dampers is also investigated.
\end{abstract}


\end{frontmatter}

\section{Introduction}
\begin{figure}[h]
\centering
\includegraphics[trim = 20mm 70mm 20mm 50mm,clip, width=17.0cm]{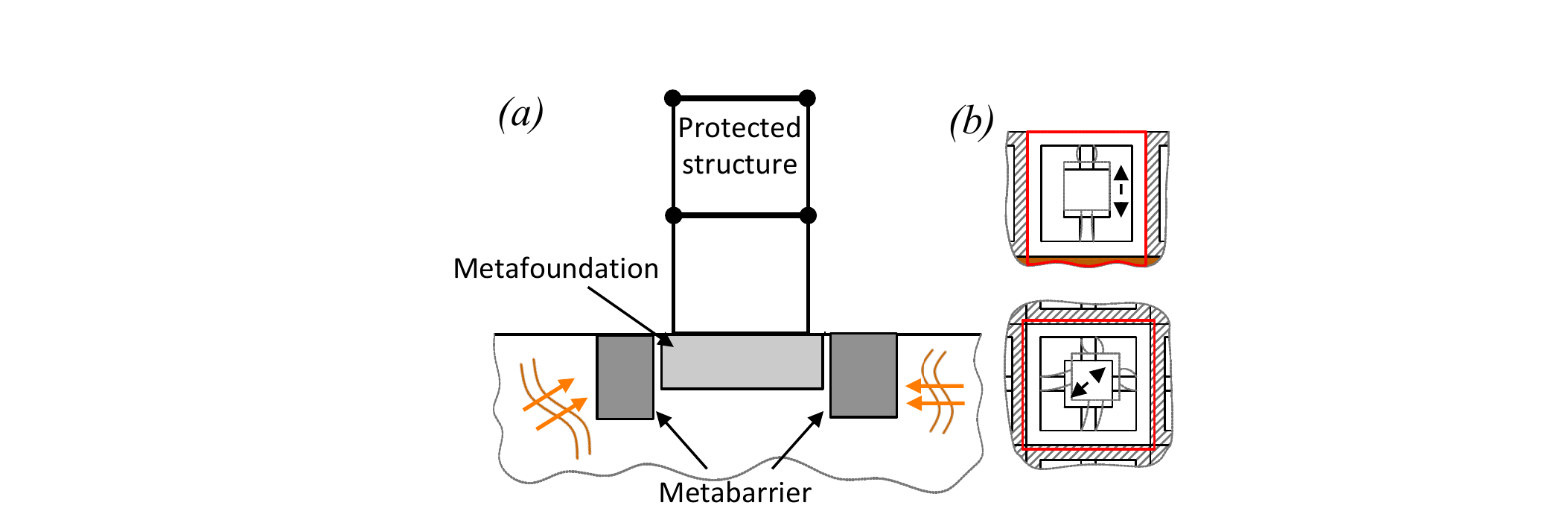}
\caption{(Color online) Different purpose and set-up of the metabarrier vs. metafoundation. The metabarrier surrounds the structure. The metafoundation also bears the weight of the superstructure. (b) Sketches of two types of resonant unit cell of resonant metamaterial characterised by a different motion polarisation.  \label{fig:fig1}}
\end{figure}

To reduce structural damage and increase safety and resilience, the engineering community is constantly looking for cutting-edge technologies to counteract the adverse effects of 
seismic waves and man-made groundborne vibrations on civil infrastructures. A substantial step forward in this direction is expected to come from the integration of so-called metamaterials in the building design. Applied mainly in the solid-state physics field, metamaterials opened up a new range of possibilities to control and shape the propagation of waves \cite{handbook_meta}. Although metamaterials are available in a variety of shape and size, all can be defined as materials featuring repeated, local (ideally very small) heterogeneities \cite{craster2012}. Thusly engineered structure influences the propagation of waves endowing metamaterials with very particular properties such as bangaps, enhanced wave guiding and absorption. Spatial arrangement of locally resonant elements \cite{matthieu,younes2011}, periodic (phononic/photonic crystals) \cite{page1} or non-periodic layouts of heterogeneities (voids, scatterers, defects) \cite{kaina_composite} inserted in a medium are by far the most common designs of metamaterials found in the literature.
In the fields of elastodynamic, and in particular seismic waves and vibrations, the most interesting model appears to be the resonant one (Fig.~\ref{fig:fig1}) although notable exceptions exist based on the well known Bragg scattering in periodic structures \cite{brillouin,kittel}. Inspired by early works on surface wave attenuation \cite{woods,sanchez_sesma} and the progresses achieved in the field of phononic/photonic crystals, Brule et al. \cite{brule} studied the effect of a periodic arrangement of pile foundations in soft soil (e.g. $v_s<$ 400 m/s) as if it was a giant soil-phononic crystal with bandgap at $\sim$ 50 Hz. A more promising variation of this study, potentially capable to attenuate low frequency seismic waves (e.g. $<$ 10 Hz), has been achieved by clamping the pile foundations to a stiff deep layer \cite{clamped_achaoui}. Interestingly, a recent work on groundborne vibrations mitigation obtained similar results \cite{ALAMO2019274}. On the resonant front, seminal studies carried out by \cite{krodel2015} in a laboratory setup and by \cite{andrea_tree,Roux2018} in a large scale experiment using trees have shown that local resonance may yield better attenuation performance than Bragg scattering and in particular achieve lower frequencies by fine tuning the resonant mass. This has been applied practically by \cite{1367-2630-18-8-083041} and \cite{Palermo:2016aa} who proposed metabarriers capable to attenuate surface seismic waves, primarily as horizontally polarised shear (SH, Love) and Rayleigh waves (e.g. Fig.~\ref{fig:fig1}).
Moving to the building itself, a large number of studies targeted the design of novel foundations. Periodic, sandwich-like slab foundations, have been studied by \cite{Xiang_2012} and \cite{babao}. Casablanca et al.\cite{finocchio2} integrated resonant elements in the slab to push the bandgap at lower frequencies while reducing the overall dimensions. It remains unclear whether these devices would offer better performance and fewer drawbacks than state-of-the-art base seismic isolators \cite{isolator,talbot} including the capacity to withstand large deformations (i.e. tens of centimetres), or offer  improved durability and serviceability. To better frame the size issue, we recall that certain metastructures relying on periodicity (Bragg scattering) \cite{brillouin} attenuate waves at the frequency when $d$=$\lambda$/2, with $\lambda$ being the wavelength and $d$ the lattice spacing. While this is fine for high frequency groundborne vibrations ($>$ 100 Hz), for low frequencies and, in particular, seismic waves, it means a periodicity $d$ of the order of tens of metres. Such a long scale variation could be better implemented within the building structure itself as shown by \cite{timoshenko} than in the soil. Therefore, if one targets the building foundations or its surrounding soil, resonant metamaterials are better suited as they support low frequency bandgaps using meter-size resonators as those presented in this study and by earlier works \cite{wenzel17, ACHAOUI201630}. Given the large dimensions, a seismic metamaterial should contain the lowest possible number of resonators (so called unit cells), to keep the cost and the total engineered area to a minimum. 

An engineering area where metamaterial applications are still largely lacking is the one dealing with the containment and mitigation of groundborne vibrations \cite{vibrations}. Because of the reduced complexity of the problem when compared to the seismic isolation one (small displacements and higher frequencies) and the technological importance of the issue, a surge in the number of studies investigating their feasibility is anticipated. 
Mostly generated by anthropic activities (e.g. railways, heavy industry and excavation), groundborne vibrations are an increasing source of problems in dense urban area, high precision laboratories and manufacturing sites \cite{takemiya2005environmental,lai}. Current technical solutions tackle the building foundations \cite{underground_vib,noise_railway,ALAMO2019274}, foresee the use of base isolator \cite{talbot} and the construction of trenches \cite{THOMPSON201645}. Seismic metamaterials could integrate those existing solutions introducing a step change in their performance.

With the exception of \cite{finocchio2}, no full-scale experiments have yet been conducted on seismic metamaterial devices and most of the work has been carried out analytically \cite[e.g.][]{eleni2,doi:10.1504/IJEIE.2016.080032,finocchio2014} or just by focusing on the single unit cell.
Typical solid state physics approaches (i.e. those applied in electromagnetic and kHz-MHz elastic metamaterials) compute metamaterial propagation properties as a function of the frequency (i.e. the dispersion curve) relying on the Bloch-Floquet conditions as if the unit cell was infinitely repeated. Whereas this only requires modeling (analytically or numerically) of a single unit cell (e.g. Fig.~\ref{fig:fig1}b)
 it produces poor results for non-periodic or finite size structures. Furthermore it cannot address a number of key engineering issues such as soil-structure interaction, the effect of spatial heterogeneities and impedance mismatches. On the other hand, full-scale 3D simulations provide deeper insights but are far more challenging for models mixing the long wavelength information (e.g. seismic waves propagating in a realistic soil) and small-scale details (e.g. the resonators and the building structure). This is particularly true for time domain simulations whose time-step is dominated by the smallest stiff element according to the CFL conditions \cite{finite_elem,FaMaPaQu97}. Nevertheless, the progress of high performance computing and state of the art elastodynamic softwares, makes 3D analysis feasible and available to seismic engineers \cite{abell,XU2019331,doi:10.1177/1094342016632596}.
Currently there is a lack of comprehensive numerical studies on seismic metamaterials \cite{wenzel17}, in particular those combining metamaterials with the superstructures. As found by Basone et al \cite{basone}, and further confirmed in our study, complex interactions between the superstructure  and the resonators may occur resulting in a dynamic reminiscent of tuned mass dampers \cite{mcnamara,tmd} and coupled resonators \cite{TOMBARI2018219}. A holistic approach that includes the structure, the metamaterial, and the surrounding soil is therefore essential.

By performing full-scale time dependent numerical simulations, this study presents a thorough validation of two potential seismic metamaterials (Fig.~\ref{fig:fig1}a) (1) the metafoundation, that both attenuates seismic waves and supports the superstructure and (2) the metabarrier, that surrounds the structure reducing the magnitude of a potentially approaching seismic waves. In both devices, a graded design is implemented by varying spatially the mass of the resonator so that a large bandwidth can be attenuated. The considered seismic scenario is that of a soft sedimentary soil (with various shear velocity) where shear waves are dominant (both surface and bulk waves) and the earthquake source is located in the far field. No faulting or permanent deformations are considered \cite{Gazetas2015,Faccioli2008}.
The study starts with a detailed analysis of the metamaterial unit cell containing the resonator. Next it considers the design of a metabarrier and a metafoundation using a different number of unit cells with spatially varying (graded) properties. Finally, using large-scale parallel simulations, the performance in term of attenuation of the spectral amplifications for both metasolutions is benchmarked considering the effect of damping in the metamaterial, the wavefield directivity, different soil properties and the size of the metamaterial.

\section{Unit cell properties and metamaterial design}
The unit cell of the metamaterial considered in this study is depicted in Fig.~\ref{fig:fig2}a. The resonator, a concrete cube-shaped mass ($E_m=30\cdot10^9$ Pa, $\nu_m=0.25$, $\rho_m=2400$ kg/m$^3$) of side $l_r=1$ m, is supported by six rubber elastic connectors ($E_l=10^6$ Pa, $\nu_l=0.44$, $\rho_l=1000$ kg/m$^3$) enclosed in a concrete hollow box ($E_b=30\cdot10^9$ Pa, $\nu_b=0.25$, $\rho_b=2400$ kg/m$^3$) of side $l_e=1.4$ m. This design supports the full complexity of 3D elastic metamaterials where resonances couple with both compressional and shear motion. Our findings are thus applicable to different unit cell designs, be it basic or more complex, that might be considered in other studies \cite{Palermo:2016aa,Dalessandro2018,wenzel17}.
A frequency domain analysis, performed on the unit cell between 0-10 Hz, using Comsol Multiphysics, reveals two peaks associated with the translational and rotational resonance (Fig.~\ref{fig:fig2}b). While the emergence of the rotational mode depends on the excitation mechanism, the translational mode is less sensitive to position and direction of the external force. In Fig.~\ref{fig:fig2}b the cell was excited laterally from the bottom side. 
It is common practice to represent the metamaterial dynamic properties through their dispersion curves, showing which modes are propagating as a function of frequency and wavenumber $k$. From the dispersion curves one can readily calculate the phase velocity as $v=\frac{\omega}{k}$ and the group velocity $v_g=\frac{d\omega}{dk} $ with $\omega $ being the angular frequency in rad/s. Dispersion curves are usually obtained by solving an eigenvalue problem on the unit cell through the application of the so-called Bloch-Floquet boundary conditions on the cell faces \cite{brillouin}. Such a boundary condition extends the unit cell as if it was infinitely periodic in the $x$,$y$ and $z$ direction. The wavenumber space $k_x$, $k_y$, and $k_z$  spans the so-called irreducible Brillouin zone \cite{brillouin} that is bounded by a $0$ to $\pi/l_e$ periodicity along the path drawn in Fig.~\ref{fig:fig2}c by the points $\Gamma$-X-M-R-$\Gamma$.  
An analytical formulation has been derived for a 1D arrangement of resonators (see Supplement material) for a single polarization. However, to precisely capture the interplay between P and S motion Fig.~\ref{fig:fig2}c show the metamaterial numerical dispersion curves for the P and S polarisations. 
The flat branches arise from the hybridization phenomena\cite{cocacola,fano} of the wide band dynamics of the background medium (a concrete block) with the narrow band response of the resonator and define the effective wave propagation along a given wave vector.
Because of symmetry, the resonant modes degenerate along the $x$,$y$ and $z$ direction and the curves overlap. Whereas the three translational resonances occurring at $\sim$6 Hz correspond to the onset of the bandgap (shaded area), the rotational modes do not yield a noticeable gap. It is worth noting that the bandgap is omnidirectional, meaning that the metamaterial affects the propagation of seismic waves regardless of direction.
The right hand side of Fig.~\ref{fig:fig2}c shows the evolution of the band diagram when changing the resonator mass. In this case, the mass has been increased to $\Bar{m}=1.6m$. The bandgap $vs.$ resonator mass plot suggests that for heavier masses there is an inverse relationship between frequency and the bandwidth. This is the key characteristic exploited by graded metamaterials \cite{krodel2015,colombi16a}.
Figure \ref{fig:fig3} focuses on the $\Gamma$-X region of the frequency-wavenumber plot. Here the P and S dispersion continuous line of the casing (a hollow concrete box) is overlaid over the numerical and analytical line of the metamaterial. The flat branch and the bandgap created by the resonance is clearly visible. To better understand the difference between the continuous concrete block, the hollow concrete casing and the unit cell with resonator Fig.~\ref{fig:fig3}b zooms on the dispersion curves of the casing and the continuous concrete block. The presence of the cavity reduces the wave velocity of the concrete ($v_p\sim3900$ m/s and $v_s\sim2300$ m/s in a continuous media). The reduction is proportional to the size of the cavity. This is a well-known property of periodic structures \cite{mouldin_waves,Khelif2004}. A curve for $v=300$ m/s is given as a proxy for a typical sedimentary soil to highlight the impedance contrast between the concrete casing and the soil.

The red lines in Fig.~\ref{fig:fig3}a represent the analytical dispersion relationship. A basic mass-in-mass model has been derived for the translational mode in an attempt to approximate the unit cell (see Supplementary material for the derivation). The eigenvalue problem of a periodic mass-in-mass systems has been solved, leading to the scalar dispersion curve of the translational mode. The values of the upper and lower gap frequencies obtained analytically coincide with the numerical ones. This simple analytical approach could be used preliminarily to work out the mass and the effective stiffness required. For a thorough analysis however, a full-scale numerical simulation should be prepared. Damping was not taken into account for the unit cell design but it will be introduced in the full-scale numerical simulations.

The technical details regarding the realization of the resonator kinematics (e.g. connector design, damping properties and bearing capacity) which allow to obtain the dynamic properties discussed are not tackled in this study. Some recent works provide more details on potential technical solutions \cite{wenzel17,basone}. We have instead carried out a static analysis and concluded that the deformation of the resonator at rest, under its own weight, is lower than 1 cm. Such a pre-stress state is not considered in the numerical model.

After presenting the unit-cell property it is time to move to the metamaterial lay out. In Fig.~\ref{fig:fig4} the unit cells are periodically arranged in the soil to design two seismic metamaterials solutions: the metabarrier and the metafoundation (respectively Fig.~\ref{fig:fig4}b and Fig.~\ref{fig:fig4}c). The main difference between the two is that the metabarrier is not required to withstand heavy vertical loads whereas the metafoundation does since the superstructure or part of it, is built on top. Both solutions are interesting concepts for the following reasons: (1) they may transfer the burden of the seismic design from the building to the foundation and the barrier, (2) they could protect more than one structure at a time and (3) they can integrate graded design enlarging their effective attenuation frequency band. Furthermore, the barrier could be scaled and frequency-tuned to mitigate groundborne vibrations, another fundamental problem in civil and mechanical engineering \cite{THOMPSON201645}. While this study is centred on horizontally polarised shear waves, these resonators could equally couple with vertically or elliptically polarised waves that are dominant in groundborne vibrations. The physics of this interaction has been thoroughly studied in previous works \cite{metaforesta,andrea_tree,Palermo:2016aa,PALERMO2018265} and applies nearly unchanged in our case.

The concept of graded metamaterials is implemented in both configurations. As discussed previously, the translational resonant frequency of each cell is tuned acting on the mass density of the resonator (Figs.~\ref{fig:fig4}b and c), which increases toward the centre of the seismic metamaterial. A decreasing distribution could lead to different wave phenomena and mode conversion \cite{colombi16a} but does not enlarge the width of the bandgap. 
The study of graded, disordered or spatially patterned metamaterials \cite{celli2019bandgap} is a novel and interesting branch. However it is not addressed in this article as it would require further analysis that may distract from the main benchmark.
On top of the graded mass distribution, a random mass error of up to 5\% has been introduced haphazardly in each resonator. This should account for the uncertainty related to the fabrication and tuning of each unit cell. Finally, we also considered a barrier without resonators (Fig.~\ref{fig:fig4}d), to rule out the effect of the concrete hollow cells on the overall attenuation produced by the metamaterial.

\section{Numerical implementation of the time domain simulation}
Full 3D numerical simulations of seismic waves are implemented using SPECFEM3D: a parallelized, time domain, and spectral element solver for elastodynamic problems widely used in the seismology and engineering community \cite{specfem_cart,Komatitsch01041998,gharti,gueguen_andrea}. The reference computational domain, depicted in Fig.~\ref{fig:fig4}a, is a 60 m depth halfspace 200$\times$100 m wide of homogeneous sedimentary material with a homogeneous shear velocity $v_s$. Numerical tests and previous work indicate that the interaction between the soil, the metamaterial and the building are dominated by the shear modulus \cite{andrea_tree}. Therefore we run simulations for different values of $v_s$ between 400 and 150 $m/s$ whereas the compressional velocity as well as the density are $v_p$ kept constant at 1000 m/s and  $\rho=$1900 kg/$m^3$ respectively. The computational 
region, apart from the free surface itself, is enclosed in perfectly matched layers (PMLs) \cite{Komatitsch_CPML} mimicking radiation to infinity and preventing unwanted reflections from the computational boundaries. These are standard procedure in elastic wave simulation allowing for long time series free of spurious reflections and essential for a thorough frequency domain analysis. A control building enclosed in the barrier/foundation is used to benchmark the metamaterial attenuation capacity. The structure fundamental mode (mainly a horizontal motion as shown in the inset in Fig.~\ref{fig:fig5}a) can be tuned numerically acting on its weight, i.e. increasing or decreasing the density of the upper hexahedral block, without altering its dimension. The resulting resonant frequency varies approximately between 3 and 10 Hz, as in flat concrete or masonry buildings.  In the case of the barrier, a 40 cm thick slab of concrete is used as a foundation to prevent rocking-like behaviour  of very soft soil and to homogenize the building motion between the barrier and foundation case.
The driving input $F$ has equal components along the $x$ and $y$ directions generating mainly SH waves. In real scenarios, SH motion is typically associated with Love waves and basin resonances as well as with vertically propagating body waves \cite{global_seism,boue}. Because of the coupling at the surface, small amplitude Rayleigh waves are also generated. Compressional waves leave the computational domain and pass into the PML at the bottom and side boundaries of the computational domain, therefore, their interaction with the structure is negligible. The position of the line source varies between the two locations (Fig.~\ref{fig:fig4}a) to include different incidence angle of the incoming waves on the structure. Some configurations consider also a SH wave with vertical incidence. In this case the source plane is located 50 m underneath the surface and it is centred on the metastructure. Even in this scenario, the forcing term acts along the $x$ and $y$ directions. The source time function is a Ricker wavelet centred at 6 Hz to sufficiently illuminate the target frequency range (see Fig.~3a). Each simulation produces a signal of approximately 3 s, enough for a detailed time-frequency analysis. 
Simulations are performed in the linear elastic (or linear viscoelastic) regime. 

Models like that in Fig.~\ref{fig:fig4}a combining the fine scale of the metamaterial and the long scale of seismic wave propagation often suffer from prohibitive computational costs if the domain is not discretised correctly. Since the purpose of the study is to create a database allowing for a sensitivity analysis over several parameters, the meshing operation is a pivotal step. Whereas the spectral element method delivers great performance in terms of computational efficiency and accuracy, its hexahedra-based mesh badly tolerates deformed elements and overly complex geometries. To keep the element size under control and avoid malformed geometries we used an adaptive strategy to create a structured grid. The mesh is generated using the commercial software Trelis 16.5 and it is streamlined using python scripting. It is therefore straightforward to implement not only the different metamaterial configurations depicted in Fig.~3b-d but also to check the impact of a different number of unit cells (in the lateral and vertical direction), the spatial mass variation within the graded metamaterial and some random variations in the material properties necessary to mimic possible fabrication defects.
For each tested design (i.e. metabarrier and metafoundation), we compute the 3D wavefield for different fundamental resonant periods of the control building to capture the frequency dependent attenuation capacity of the metamaterial. The control building is designed to oscillate in a horizontal manner in either $x$ or $y$ direction (see the inset in Fig.~\ref{fig:fig5}a). Note that without the control structure the characterisation of the metamaterial effect would have been more difficult. Fig.~\ref{fig:fig5}b shows the seismic traces before, inside and after the barrier. Differences are marginal and can only be seen after a detailed analysis (see Supplementary Materials).  The building roof drift as a function of time is depicted in Fig.~\ref{fig:fig5}c for one protected configuration (e.g. Fig.~\ref{fig:fig4}b or Fig.~\ref{fig:fig4}c ) and as a point of  reference, unprotected one (Fig. ~\ref{fig:fig4}d). After Fourier-transforming the signals, the horizontal resonant peaks are isolated enough to prevent misidentifications as depicted Fig.~\ref{fig:fig5}d.
The amplitude difference at the building resonance between protected and reference configuration is estimated as shown in Fig.~\ref{fig:fig5}d and  plotted in Fig.~\ref{fig:fig5}f for control structures with different fundamental mode. Such a plot indicates the amplification reduction and bandwidth, roughly corresponding to the natural frequencies of the resonators, as pointed out by Fig.~\ref{fig:fig5}e.
\section{Results}
The amplification reduction curves plotted in the figures \ref{fig:fig7} to \ref{fig:fig10} are the result of an average over the wavefield computed for the two line sources depicted in Fig.~\ref{fig:fig4}a and over both the $u_x$ and $u_y$ components for the surface source setup. In the vertical incidence case instead, we average only over $u_x$ and $u_y$ because a single source configuration is considered.
Therefore, each value (the dots in the amplification reduction plots) requires four simulations with the same resonant frequency of the control building for the surface source, and only two in the vertical incidence one. 
Given the large number of points necessary to obtain meaningful amplification reduction curves, one can easily appreciate the high number of simulations computed for this analysis. 
Finally we recall that the metamaterial simulations include a random 5\% variation of the mass of each resonator to account for defects. 

\subsection{Influence of the concrete cell structure embedded in the soil}
The first test in Fig.~\ref{fig:fig6}a measures the impact of the concrete cell structure with no resonators inside the casing, against the homogeneous soil case where no structure other than the building and the slab foundation is present. The tested configurations, as in the remaining of the paper, consider three different background soil velocities $v_s$ to identify distinct behaviours of the barrier in soft and firmer soils. In the target frequency range, the shear wavelengths vary approx. between 20 to 100 m depending on $v_s$ and frequency.  The dispersion curves in Fig.~\ref{fig:fig3}b showed a casing much stiffer than the soil  (a steeper dispersion curve means higher wavespeed). However, without resonators no bandgap is generated in the 0 to 10 Hz band (although at higher frequency Bragg interference may occur). Accordingly, the ratios in Fig.~\ref{fig:fig6}a for the barrier indicates that the impact of the concrete structure, if any, is negligible when the soil is sufficiently firm. For low $v_s$, the soft-soil vs. stiff-barrier impedance mismatch plays a stronger role producing an amplification reduction $>$ 0. Value $<$ 0 might be associated to weak resonances inside the soil cavity created by the barrier when part of the energy leaks through. s
Moving to the metafoundation in Fig.~\ref{fig:fig6}b, a stronger imprint on the amplification reduction compared to the case of the barrier is observed. Further analyses suggest that soft soils (e.g. 150 m/s and lower) increase the rigid motion of the slab foundation in the reference model. Instead, the larger footprint of metafoundation prevents this phenomenon resulting in a 15-20 \% reduction of the peak amplification as indicated by the blue line in Fig.~\ref{fig:fig6}b. The vertical incidence case leads to similar conclusions and it is therefore not shown here.

\subsection{Metabarriers of different lateral widths}
Having assessed the impact of the hollow concrete cell structure on the wavefield, we begin the analysis of the actual resonant seismic metamaterial. We commence by exciting a barrier with a SH surface source as in the previous case. Figure \ref{fig:fig7}a shows the amplification reduction induced by a barrier featuring only two lateral resonator layers (hence a thickness of 2.8 m), while Figs.~\ref{fig:fig7}b and \ref{fig:fig7}c four (5.6 m) and six (8.4 m) lateral layers, respectively. The number of vertical layers is kept fix to five for now (depth equal to 7 m).  The shaded grey area on the reduction plots highlights the theoretical bandgap considering the cumulative effect of the graded masses (Fig.~\ref{fig:fig2}). For instance, in Fig.~\ref{fig:fig7}a, the resonators of the outer layer in the $x$ or $y$ direction have a mass of $m$, while the innermost ones have a mass of $1.1m$. In the six layers case (Fig.~\ref{fig:fig7}c), the innermost ones reach $1.5m$. In the 2 layers case, the reduction reaches  35-45\%  however due to  limited grading, the bandwidth is very narrow (1.5-2Hz at 25 \% reduction, see Fig.~\ref{fig:fig7}d). On the other hand, when a proper graded design is implemented, a continuous and broadband reduction can be achieved. This can be appreciated by looking at the six layers case depicted in Fig.~\ref{fig:fig7}c where the ratio top the 50\% over nearly 4 Hz. Performance indicators such as maximum reduction and bandgap size as a function of the number of lateral layers are collected in Fig.~\ref{fig:fig7}d. The maximum amplification reduction quickly reaches a steady value, approx. between 45\% and 55\% depending on the soil type. The bandwidth instead is directly proportional to the number of layers owing to the graded configuration. 
In all instances the mass of the resonators increases linearly towards the interior of the barrier. Reversing the mass distribution does not lead to any noticeable difference. 
The influence of the number of vertical layers is discussed later when using a vertical incidence wavefield.

\subsection{Metafoundations and the role of viscous damping}

The next batch of simulations continues with a surface wave excitation but now addressing the metafoundation (Fig.~\ref{fig:fig4}c) made of 16 $\times$ 16 unit cells along $x$ and $y$ and 5 along $z$. To allow for a fair comparison, the lateral extension (22.4 m) is identical to the metabarrier in Fig.~\ref{fig:fig7}b. Also the grading mimics the one in Fig.~\ref{fig:fig7}b but here the masses vary from $m$ to 1.3$m$ in the 4 outermost layers while the innermost 8 are kept constant at 1.3$m$. 
The large number of resonators integrated in the metafoundation and the continuity with the superstructure deliver higher reduction ratios compared to the metabarrier but the bandwidth remains very similar to that in Fig.~\ref{fig:fig7}b.

However, a detailed inspection of the building response spectra and the associated time series (Figs.~\ref{fig:fig8}c and d) with and without resonators reveals that building-resonators coupling affects the performance by inducing a detuning of the dynamic response, i.e. a shift of resonance of the building (Fig.~\ref{fig:fig8}d). This phenomenon, recorded in all tested metafoundation cases, appears particularly detrimental when the building resonance is located inside the bandgap as, for example, in Fig.~\ref{fig:fig8}d. The displacement plot in Fig.~\ref{fig:fig8}c shows the onset of the detuning in the coda of the signal (blue vs. red line). 
Amplification before the bandgap and detuning are the hallmarks of many resonant metamaterials and previous studies have already reported on this \cite{doi:10.1504/IJEIE.2016.080032,wenzel17}. In our case the absence of strong amplification is likely due to the presence of the graded design. 
Interestingly, by accounting for a small value of damping (2\%) in the resonators (namely in the rubber connectors), which in the previous simulations have been considered purely elastic, the response is smoothed out (Fig.~\ref{fig:fig8}c) and the reduction ratio increased (Fig.~\ref{fig:fig8}b).

Overall the metafoundation behaviour is reminiscent of a multi-tuned mass damper (TMD) where the introduction of appropriate mass distributions and damping reduce mistuning problems and smooth out the response \cite{IGUSA1994491}. An optimisation (not implemented here) of the damping value and the mass distribution could yield even higher rates of amplification reduction.
In the metabarrier case this behaviour is not reported because the presence of the soil region between the building and the barrier (Fig.~\ref{fig:fig5}c), and the consequent impedance mismatch, decouples the resonators from the superstructure.

In terms of performance, the highest reduction rate is achieved in soft soil. However the difference between firm and soft is slightly reduced compared with the metabarrier. A stronger reduction peak is localised in a narrow region roughly corresponding to the resonance frequency of the 1.3$m$ unit cell characterising the innermost 8 layers. Finally we note that only the introduction of damping unlocks the potential of the metafoundation allowing superior reduction rates $\sim70\%$ (Fig.~\ref{fig:fig8}b).

\subsection{Vertical incidence SH waves}

We now move on to the vertical incidence case depicted in Fig.~\ref{fig:fig9}. The structure is excited by a SH wave generated underneath (Fig.~\ref{fig:fig4}a). We first take advantage of this last source layout to evaluate the impact of the number of unit cell in the $z$ direction on the amplification reduction. The summary plot in Fig.~\ref{fig:fig9a}, computed for a metafoundation and a metabarrier of lateral extension as in Fig.~\ref{fig:fig9}, shows the maximum reduction for an increasing number of vertical layers. In the metafoundation case, each additional layer between 1 and 4 strongly improves the amplification reduction while the $5^{th}$ layer's contribution is limited. The results for the metafoundation are far less impressive: the reduction ratio grows linearly and the maximum reduction is less than half of the one achieved by the metafoundation and much lower than the lateral incidence case (Fig.~\ref{fig:fig7}b). Supported by the plots in Fig.~\ref{fig:fig9}, we now investigate in detail the possible cause of this discrepancy in performance.

Figures \ref{fig:fig9}b and d respectively, compare metabarrier and metafoundation with 2 layers of resonators in the vertical direction. The metafoundation shows a significant, though narrow, reduction level in line with the results reported in a previous study on a 2 layers metafoundation \cite{wenzel17}.
When comparing the metabarrier in Fig.~\ref{fig:fig9}a with Fig.~\ref{fig:fig7}b, a strong degradation of the reduction performance is visible in spite of the identical number of lateral layers and graded profile characterising both designs. Furthermore, in Fig.~\ref{fig:fig7}b the gap between stiffest and softest soil curve was rather constant throughout the frequency range while in Figs.~\ref{fig:fig9}a, and \ref{fig:fig9}b in particular, the green and blue soft soil curves (300 and 150 m/s respectively) overlap at high frequencies. In the previous case of lateral incidence and 5 vertical layers (Fig.~\ref{fig:fig7}b  vs. \ref{fig:fig8}b) the difference between metabarrier and metafoundation was less noticeable.  
This indicates a strong dependence of the barrier mitigation performance on the wavelength, the wave incidence direction and the number of vertical layers. In particular, it emerges that for a shorter wavelength (e.g. soft soil and high frequencies) and vertical incidence, energy is easily channelled through what should instead be the protected inner region ($\sim 14 \times 14$ m). The phenomena is exacerbated when the barrier is shallow (e.g. 1-2 vertical layers).

A final example of leakage through the barrier with vertical incidence is given in Fig.~\ref{fig:fig10}. Here the metabarrier is located at a greater distance from the building and the attenuation capacity is close to zero with vertical incidence. In the lateral case, the amplification reduction shows a behaviour not too distinct from the configuration in Fig.~\ref{fig:fig7}a that also shares an equal number of layers and mass distribution. 
In the light of these results the metabarrier might present  a viable solution for containing groundborne vibrations, usually confined at the surface. 
For the seismic use, where energy is characterised by a very heterogeneous azimuthal distribution, it might not be an ideal solution and further research is required.

\section{Discussion and conclusion}

\begin{figure}[h]
\centering
\includegraphics[trim = 0mm 55mm 0mm 60mm,clip, width=17.0cm]{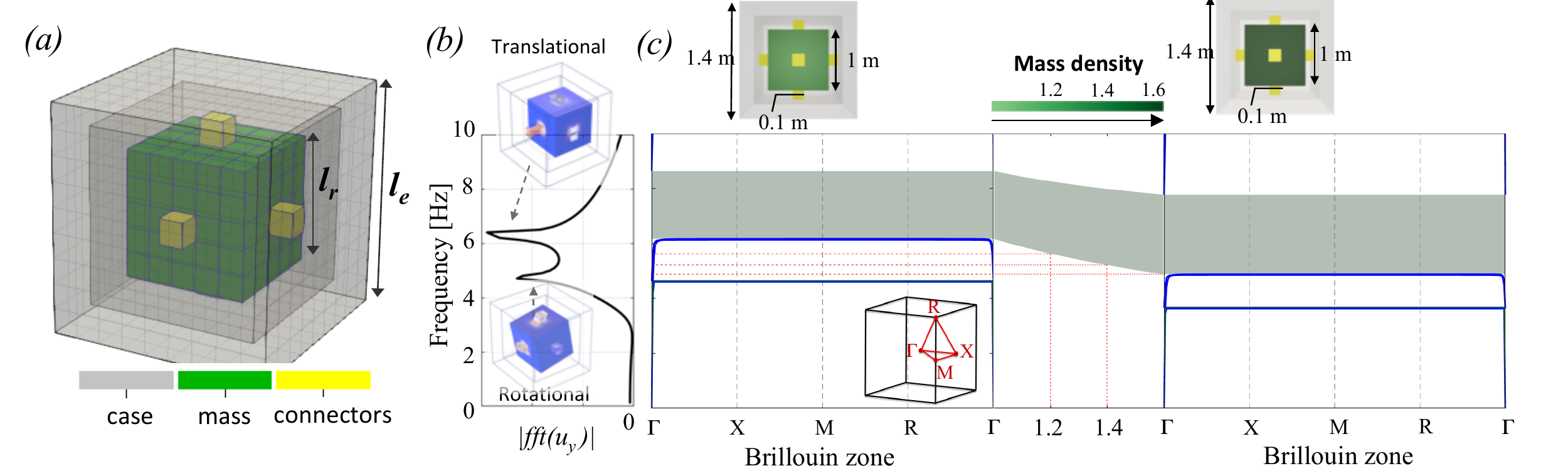}
\caption{(Color online) (a) Detailed view of the unit cell and its component constituting the metamaterial. The hollow case and the resonant mass are in concrete while the connectors in reinforced rubber. (b) The frequency response function of the unit cell for a uni-axial forcing direction and the associated modal shape. (c) The dispersion curve along the irreducible 3D Brillouin zone (according to the standard $\Gamma-X-M-R-\Gamma$ representation \cite{kittel}) for two different resonant mass weights. The shaded zone highlights the bandgap region for the given cell type.   \label{fig:fig2}}
\end{figure}

\begin{figure}[h]
\centering
\includegraphics[trim = 20mm 70mm 20mm 60mm,clip, width=17.0cm]{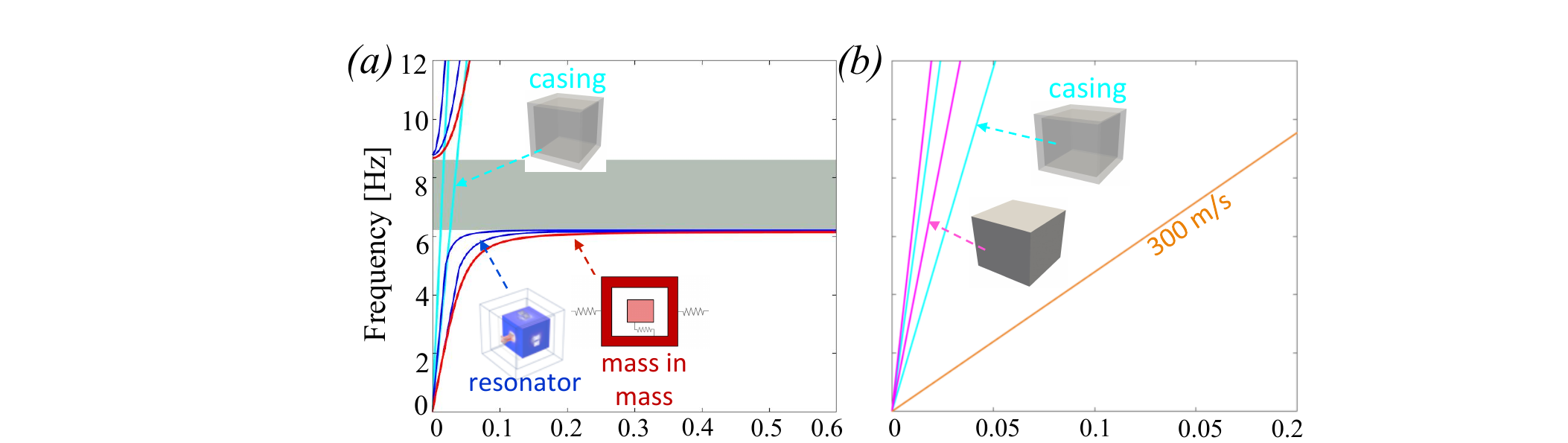}
\caption{(Color online) (a) A section of the dispersion curve in Fig.~\ref{fig:fig2} along the $\Gamma-M$ segment shows in red the analytical dispersion curve of the metamaterial in blue the numerical one and in light blue those of the casing (without the resonator) also computed numerically. The numerical dispersion curve are characterise by a P and a S mode while the analytical model has been computed for an S-like only. (b) A zoomed portion of (a) compares the P and S dispersion curve for a full concrete block against the hollow casing one. A line at $v$=300 m/s is marked for reference. \label{fig:fig3}}
\end{figure}

\begin{figure}[h]
\centering
\includegraphics[trim = 20mm 20mm 15mm 10mm,clip, width=17.0cm]{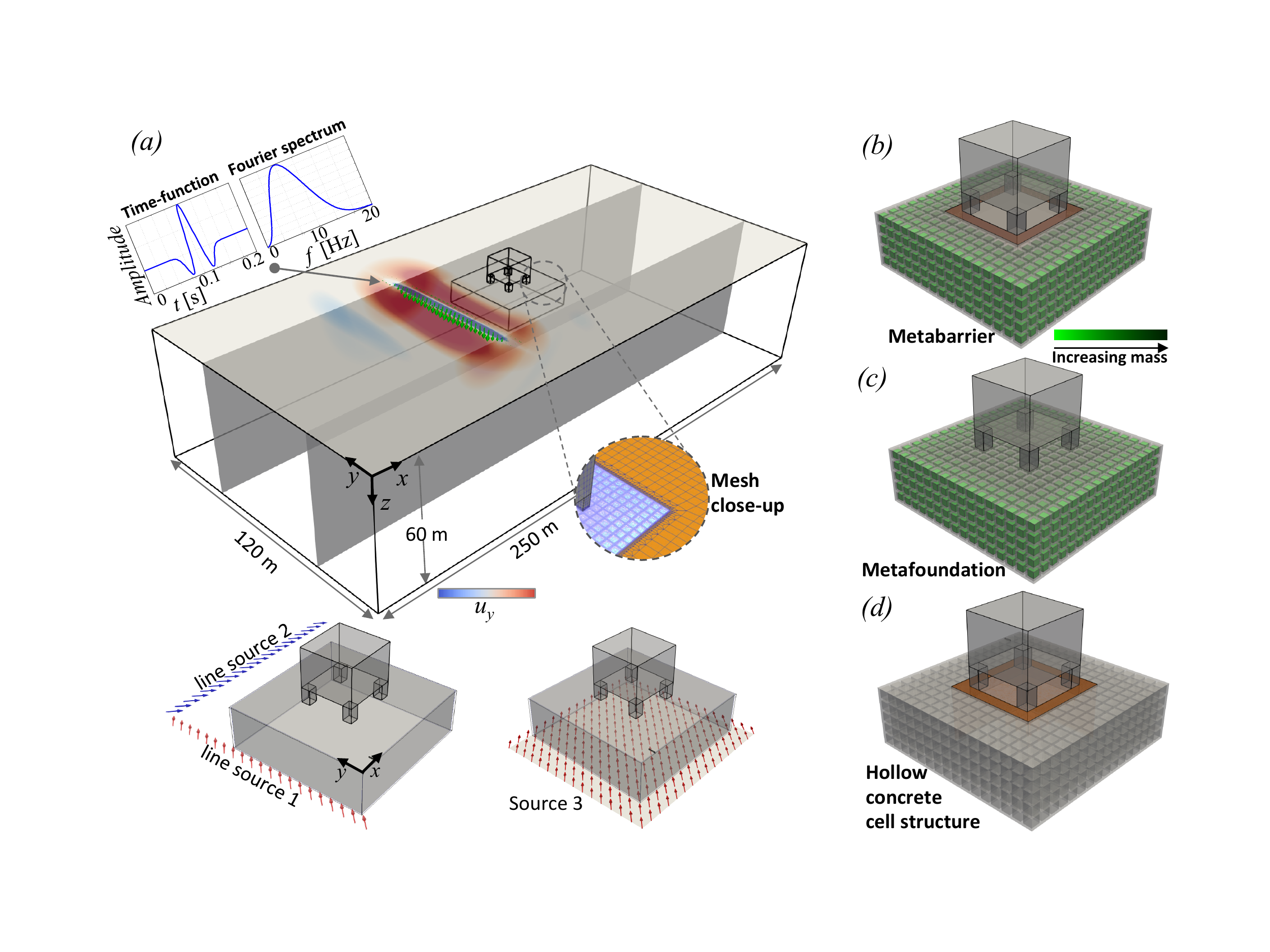}
\caption{(Color online) Simulation set-up. (a) The computational domain characterised by soil, building and metamaterial shows 3 different source locations and layout (surface lateral and bottom incidence) for a excitation driven in time by a Ricker function (inset). The arrows point toward the actual force direction at the onset time. The inset zooms on a detail of the fine mesh structure at the boundary between the metamaterial and the homogeneous soil. The colormap shows the normalised displacement field along $y$. (b-d) Different metamaterial configurations used for the simulations. \label{fig:fig4}}
\end{figure}

\begin{figure}[h]
\centering
\includegraphics[trim = 20mm 45mm 20mm 40mm,clip, width=17.0cm]{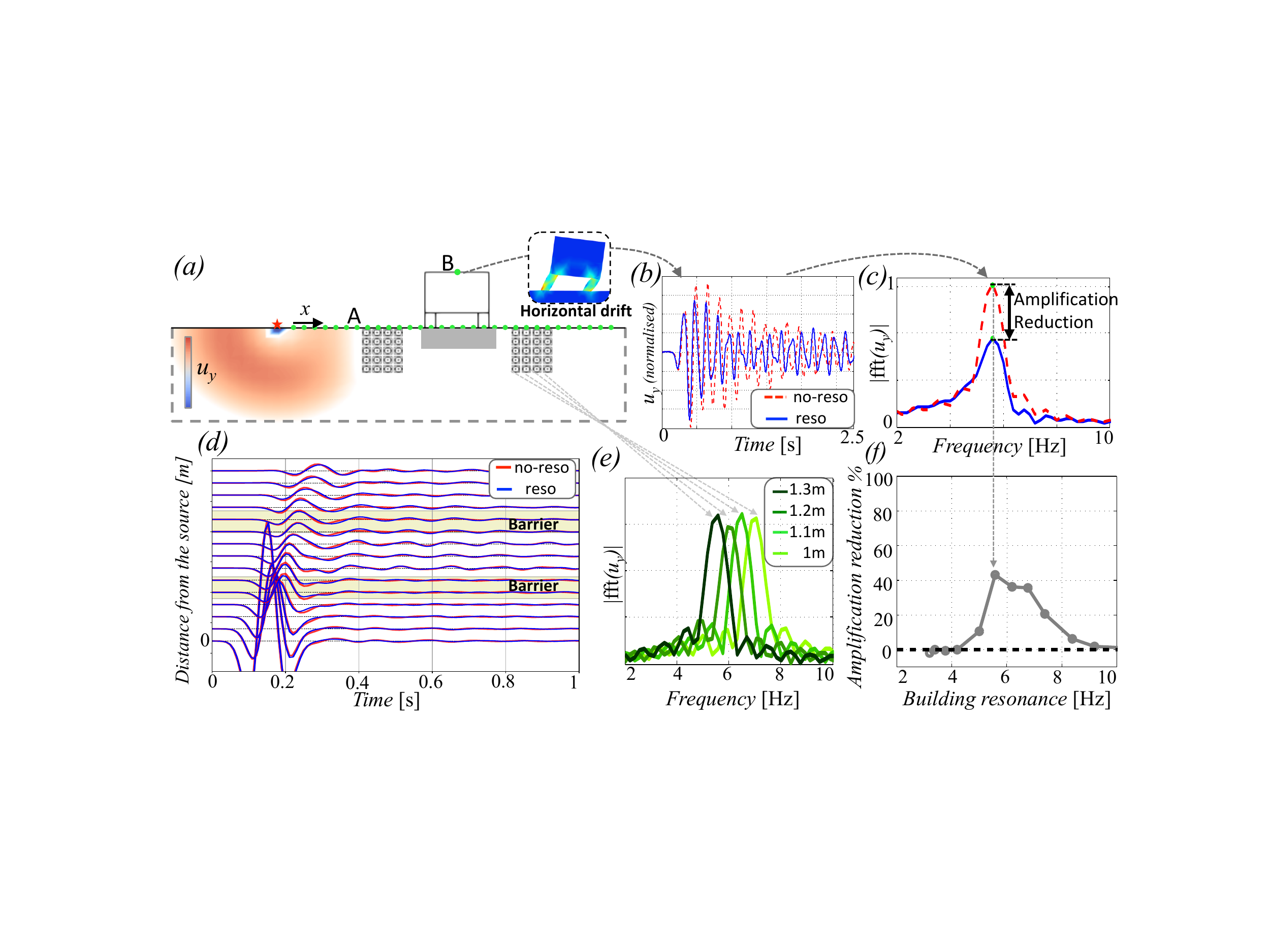}
\caption{(Color online) (a) Cut through (mid-plane) the 3D wavefield shortly after the onset time showing the control building, the slab foundation and the metabarrier. The inset shows where deformations are located in the control building at the fundamental horizontal mode. (b) Displacement $u_y$ recorded at the top of the structure (point B) for the metabarrier (reso) and the concrete hollow cell (no-reso) in Fig.~\ref{fig:fig4}d. (c) Fourier spectra for the traces in (b). The difference (reduction) is highlighted. (d) Seismic section recorded at the surface along the mid-plane for the reso vs. no-reso case. The barrier position is highlighted. (d) Fourier spectrum of four resonators located along the barrier. The colorcode represent the mass (e.g. Fig.~\ref{fig:fig2}b). (e) Amplification reduction values as shown in (c) for different resonant frequencies of the building. Each dot corresponds to a different resonant frequency and hence to different simulations. \label{fig:fig5}}
\end{figure}

The recent rise in the number of studies putting forward  metamaterial solutions for seismic waves mitigation is testimony to the fact that metamaterials are no longer regarded as a purely theoretical concept. In this study, after assuming that one can practically engineer the unit cell kinematics necessary to achieve the dynamic outlined in Figs.~\ref{fig:fig2} and \ref{fig:fig3}, we evaluate the performance of two metamaterial designs, the metabarrier and the metafoundation, for seismic wave mitigation. Instead of the usual analysis based on the unit cell dispersion curves we adopt an holistic approach based on full 3D time dependent simulations including the soil, the metamaterial, a complex wavefield and a superstructure to be protected with a realistic dynamic response. While the work targets frequencies $<$ 10 Hz and shear waves these metasolutions are  easily scalable and can be adapted for groundborne vibrations containment at frequencies $>$ 20 Hz for both shear and Rayleigh-like polarizations. 

\begin{figure}[h]
\centering
\includegraphics[trim = 20mm 50mm 20mm 50mm,clip, width=17.0cm]{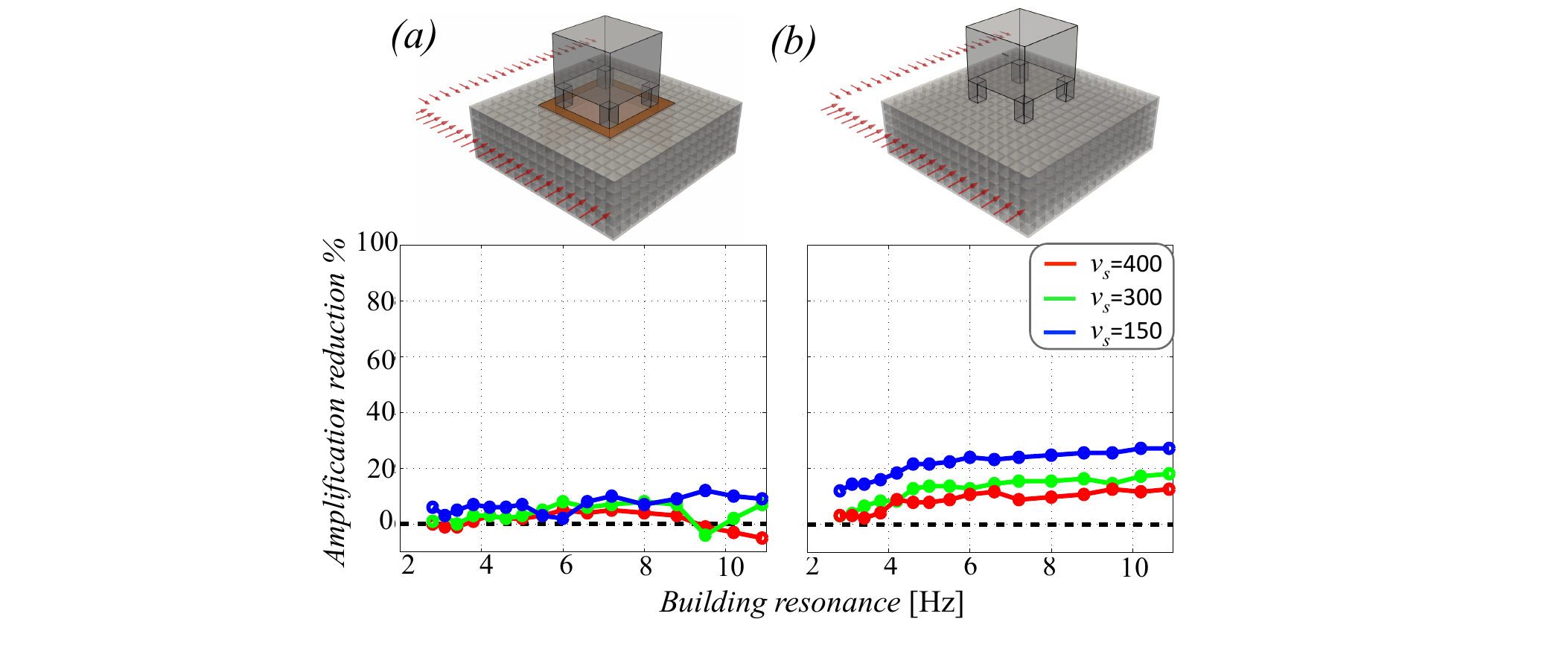}
\caption{(Color online) Reduction ratios computed as in Fig.~\ref{fig:fig5} between the homogeneous soil model and the concrete cell structure without the resonators. (a) Barrier without resonators vs. homogeneous soil model. (b) Foundation without resonators vs. homogeneous soil model. The red, green and blue lines are calculated for different background soil shear velocity. The arrows in the figure depict the incidence direction of the surface waves with respect to the structure. \label{fig:fig6}}
\end{figure}

\begin{figure}[h]
\centering
\includegraphics[trim = 10mm 50mm 10mm 30mm,clip, width=17.0cm]{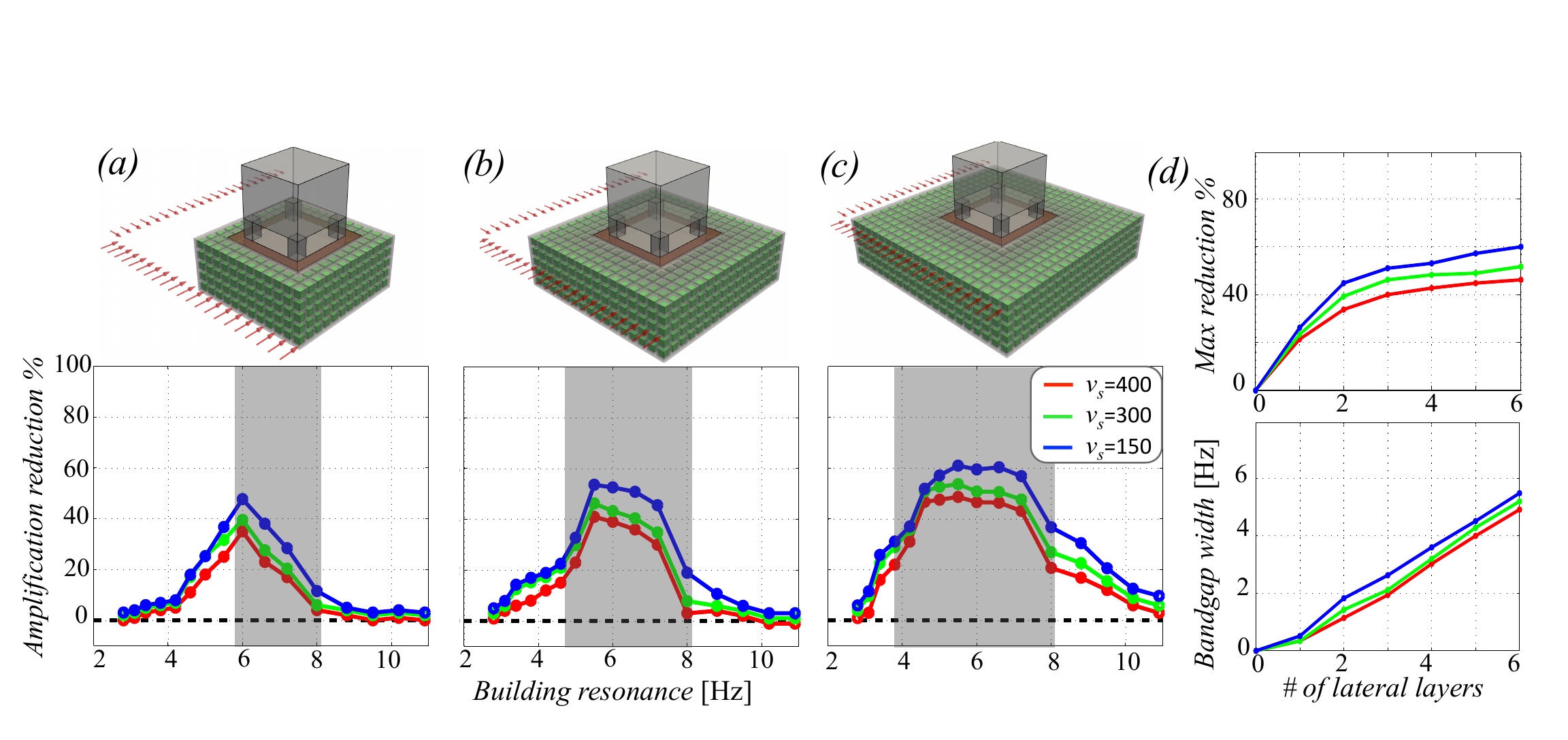}
\caption{(Color online) Same as Fig.~\ref{fig:fig6} but for the barrier concrete cell structure vs. the actual metabarrier with resonators. (a) Metabarrier made of 2 lateral unit cells, (b) 4 lateral unit cells, (c) 6 later unit cells. The number of cells in the vertical direction is kept constant to 5. The shaded area shows the predicted bandgap region extracted from the dispersion curves in Fig~\ref{fig:fig2} depending on the mass range of the resonators. (d) Summary plots showing the maximum reduction and the bandgap width (calculated for a minimum of 25 \% reduction) as a function of the number of lateral layers. \label{fig:fig7}}
\end{figure}

\begin{figure}[h]
\centering
\includegraphics[trim = 20mm 50mm 15mm 30mm,clip, width=17.0cm]{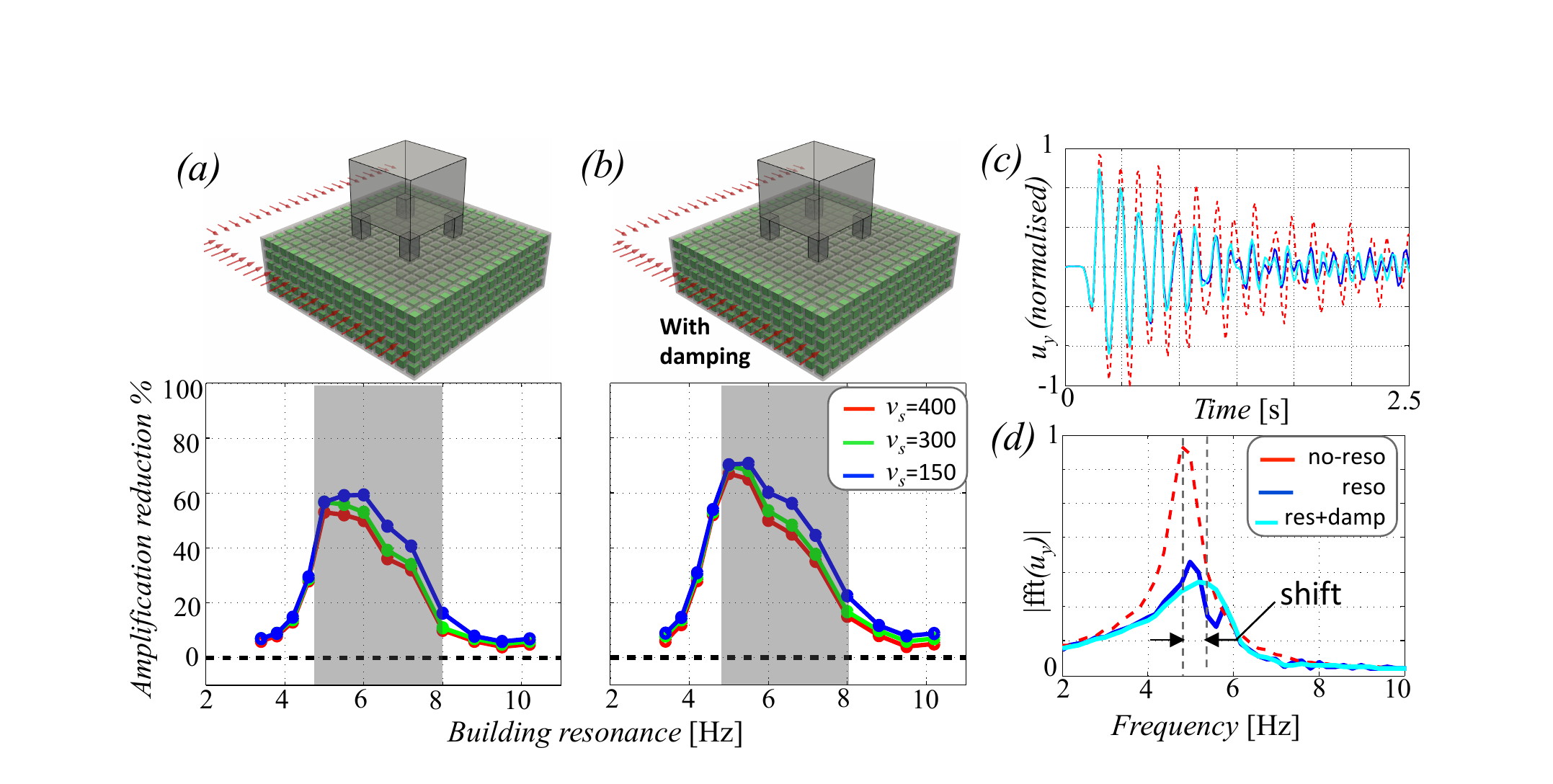}
\caption{(Color online) Same as Fig. \ref{fig:fig7} but for a metafoundation made of 16 $\times$ 16 lateral unit cells and 5 along $z$. (a) Reduction values resulting from a simulation without damping (as in previous figures). (b) Numerical simulations with a 2\% damping value in the resonator connectors. (c) Displacement $u_y$ recorded at the top of the structure (point B) and their Fourier spectra (d) for the metafoundation (reso) the concrete hollow cell (no-reso) and the metafoundation with damping (res+damp).. \label{fig:fig8} }
\end{figure}

\begin{figure}[h]
\centering
\includegraphics[trim = 20mm 40mm 20mm 30mm,clip, width=17.0cm]{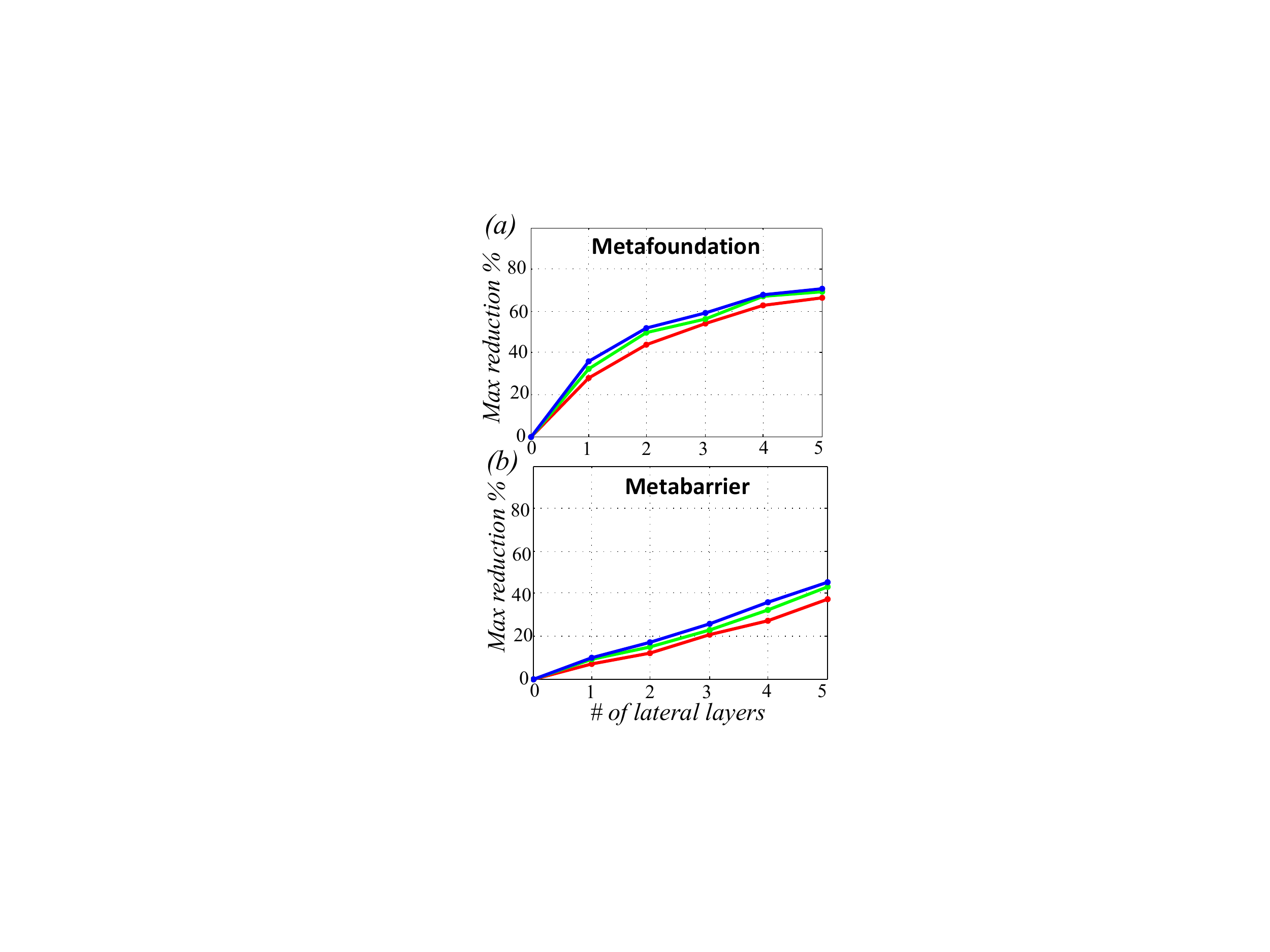}
\caption{(Color online) Summery plots showing the maximum amplitude reduction as a function of the number of vertical layers (1 to 5) for (a) the metafoundation and (b) the metabarrier calculated for a source with vertical incidence.\label{fig:fig9a}}
\end{figure}

\begin{figure}[h]
\centering
\includegraphics[trim = 10mm 50mm 10mm 30mm,clip, width=17.0cm]{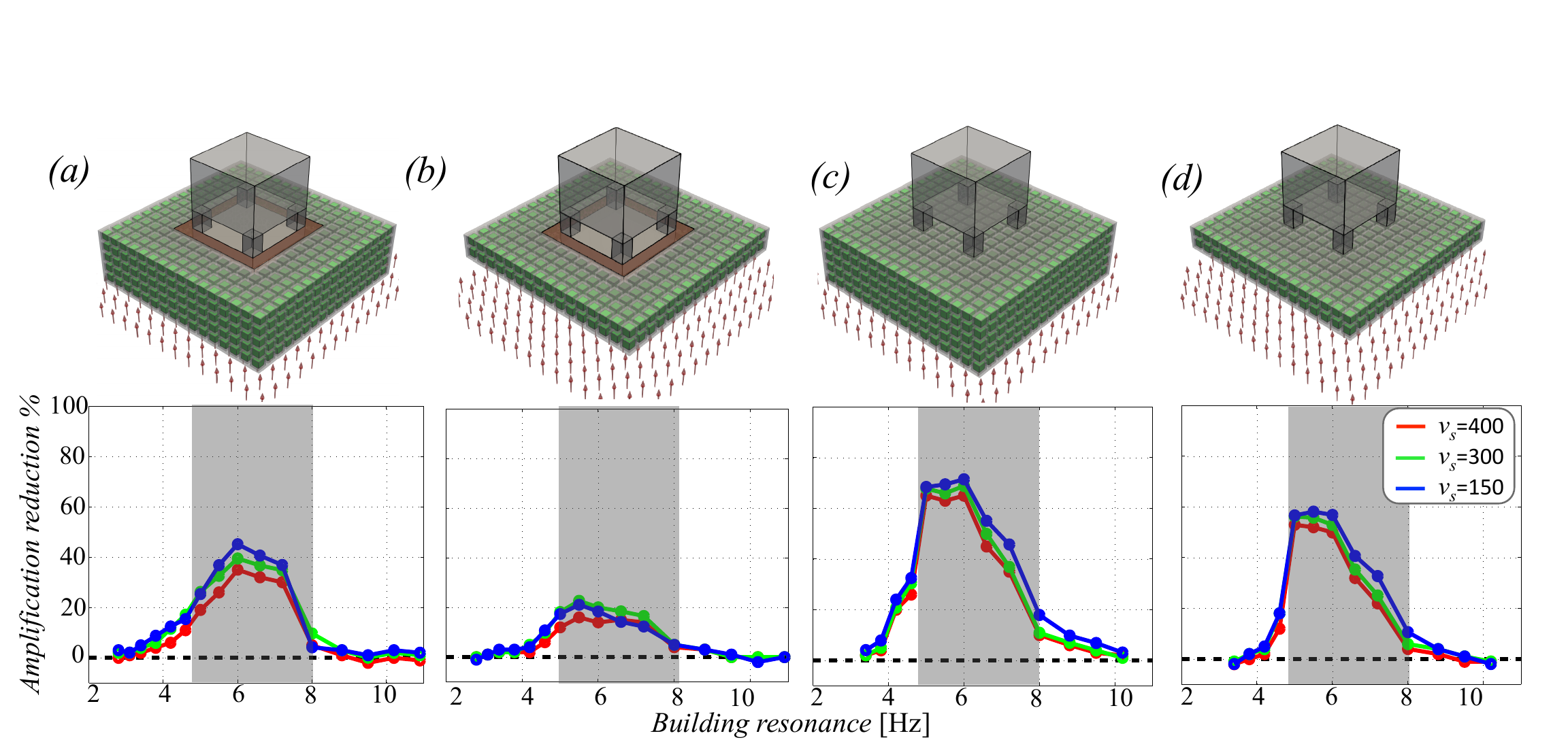}
\caption{(Color online) Same as Fig. \ref{fig:fig7} but for different metabarrier and metamaterial configurations with a vertically incident SH wavefield. (a) Metabarrier with 4 lateral unit cells and 5 along $z$. (b) Same as (a) but with only 2 cells along $z$. (c) Metafoundation as in Fig.\ref{fig:fig8} with damping in the resonators. (d) Metafoundation with 2 unit cells along $z$.\label{fig:fig9}}
\end{figure}

Despite the somewhat high computational cost of 3D simulations, we have bench tested a wealth of design parameters: the size and the number of resonators necessary to achieve protection, the benefit brought by the so-called graded mass distribution, the impact of the soil stiffness, and the source incidence direction. The results obtained could have not been inferred by looking solely at the dispersion curves.
For example, the coupling between metafoundation and superstructure reduces the effectiveness of the solution if no damping is added in the resonator's connectors. This behaviour is reminiscent of a multi-tuned mass damper \cite{IGUSA1994491} where one should tune not only the period but also the damping level and the mass distribution. In the case of the barrier, the coupling is prevented by the impedance mismatch between the barrier and the soil. In all instances considered, the analysis shows a direct correlation between the number of resonator layers and the mitigation capacity. With the exception of metabarriers with vertical incidence, each additional layer of resonators increases the mitigation capacity of the metasolution from 5 to 20\%. The graded design yield reduction ratios up to 50-60\% with a 2-3 Hz bandwidth for the metabarrier while the metafoundation reaches a remarkable 70 \%. 
The soil type is another important parameter: Softer soils yield 10 to 15\% more attenuation when compared to stiffer ones. This may be caused by the so-called soil-structure interaction that is usually enhanced by soft soils.
Contrary to other studies, our findings do not support any solution based on non-resonant soil inclusions or cavity (e.g. Fig. \ref{fig:fig6}). In the targeted frequency range, their effect, if any, amount to a slight modification of the soil effective properties without providing appropriate mitigation.
Further improvements in terms of mitigation bandwidth could be introduced by adopting non-linear connectors or by optimising damping and mass distribution. However, we expect the main conclusions of this study to remain meaningful for any type of metabarrier or metafoundation.

\begin{figure}[h]
\centering
\includegraphics[trim = 20mm 45mm 20mm 40mm,clip, width=17.0cm]{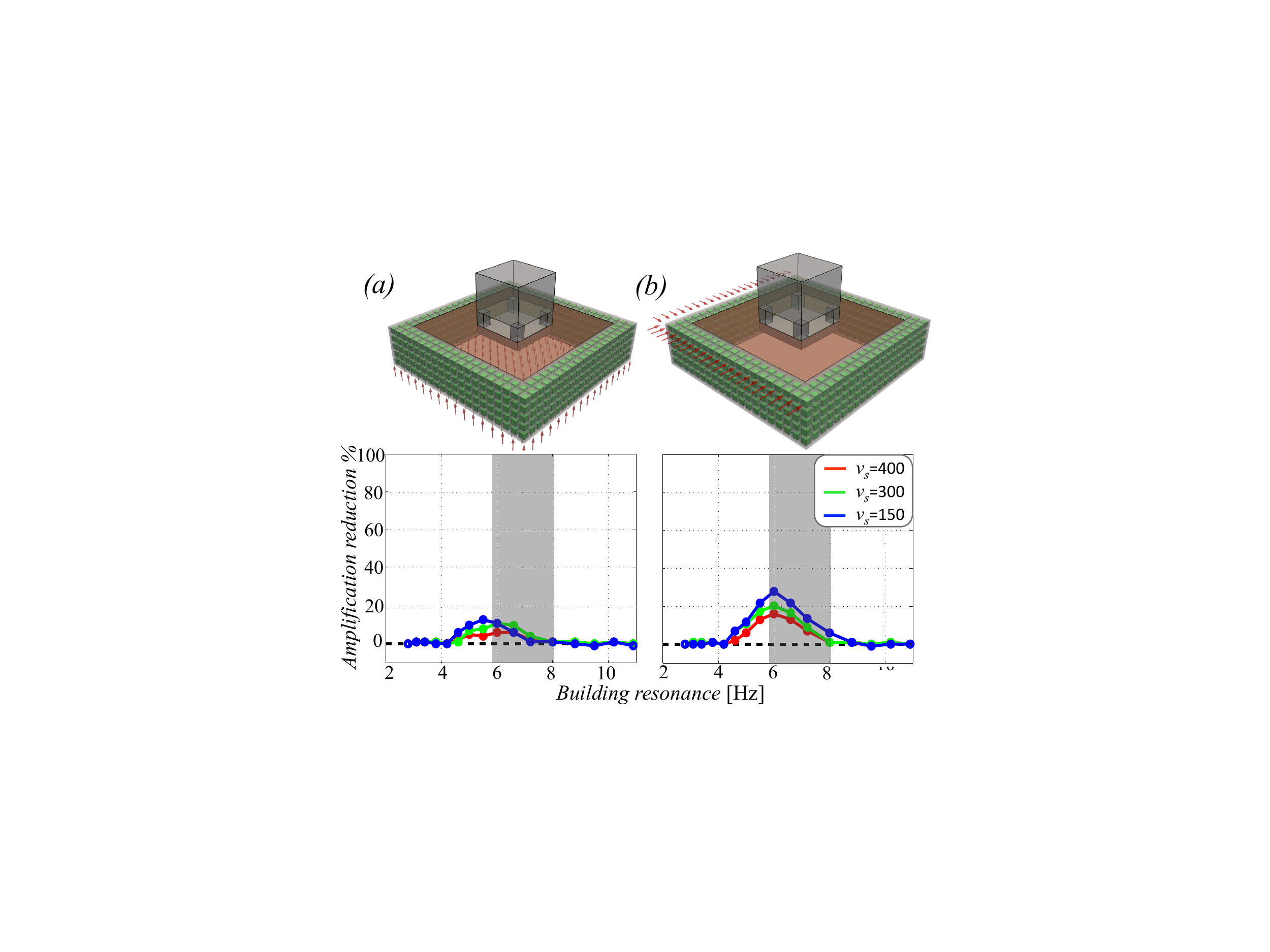}
\caption{(Color online) Amplification reduction plot as in Fig.~\ref{fig:fig7} but for a metabarrier located far away from the structure. (a) Metabarrier with 2 lateral unit cells and excited with a vertically incident SH wavefield. (b) Same as (a) but for a surface wave excitation. \label{fig:fig10}}
\end{figure}

In spite of the promising results, the conundrum of seismic protection with metamaterials remains largely unresolved.
Notwithstanding the cost of these solutions, which might prevent any practical applications, many other compelling questions arise: In the case of the barrier for instance this study is limited to a simple scenario where the distance between the metamaterial and the structure is small. What would happen if the barrier is located at distances larger than in what is shown in Fig.~\ref{fig:fig10}? Could this solution be used to protect more than one structure, or a much larger compound? Concerning the metafoundation, issues such as the vertical bearing capacity, the size required to mitigate vibration in the 1-2 Hz range and a benchmark against existing solution based on base isolators should be thoroughly considered.  
On the resonator front we have considered a displacement $<$ 10 cm although it is known that strong ground motion can easily exceed such values. Reliability and serviceability of the kinematics are also relevant issues that have been so far overlooked. 
In both metabarrier and foundation case, the total mass of the resonator should also be a design parameter. Here the total mass of the resonators is of the order of the mass of the superstructure. The adoption of smaller masses and a more deformable kinematic, possibly nonlinear, should be considered. 
Finally geomechanical soil parameters such as porosity, lateral heterogeneities, plasticity and liquefaction have not been considered but their role could be crucial.
This points out another fundamental problem: a new generation of numerical simulations is necessary for future works. In particular it is high time to develop software integrating full geomechanical and structural modelling capable to run on large parallel computers. This will allow a truly holistic, performance-based design of seismic metamaterials. 
\\\\
\textbf{Acknowledgement}
AC thanks the SNSF for their support through the Ambizione fellowship PZ00P2-174009. RZ acknowledge support from ETH research Grant 49-17-1.  
AC is particularly grateful to Ms. Alicja Crome for her accurate proofreading work.
   
%

\end{document}